\documentclass[11pt,a4paper]{article}
\usepackage{jheppub} 
\usepackage{epstopdf}
\usepackage{float}

\graphicspath{{figures/}}

\begin{document}

\title{\bf Improving the hierarchy sensitivity of ICAL using neural network}
\author[1]{Ali Ajmi,}
\author[2]{Abhish Dev,}
\author[1]{Mohammad Nizam,}
\author[2]{Nitish Nayak,}
\author[2]{and S. Uma Sankar}
\affiliation[1]{Homi Bhabha National Institute, Anushaktinagar, Mumbai 400 094, India}
\affiliation[2]{Department of Physics,  Indian Institute of Technology Bombay, Mumbai 400 076, India}

\emailAdd{aliajmi@tifr.res.in}
\emailAdd{abhishdev92@gmail.com}
\emailAdd{mnizami132@gmail.com}
\emailAdd{felarof.nayak@gmail.com}
\emailAdd{uma@phy.iitb.ac.in}

 \begin{abstract}{
  Atmospheric neutrino experiments can determine the neutrino mass hierarchy for any value of $\delta_{CP}$.
  The Iron Calorimeter (ICAL) detector at the India-based Neutrino Observatory can distinguish between the
  charged current interactions of $\nu_\mu$ and $\bar{\nu}_\mu$ by determining the charge of the produced 
  muon. Hence it is particularly well suited to determine the hierarchy. The hierarchy signature is more
  prominent in neutrinos with energy of a few GeV and with pathlength of a few thousand kilometers, {\it i.e.}
  neutrinos whose direction is not close to horizontal. We use adaptive neural networks to identify such 
  events with good efficiency and good purity. The hierarchy sensitivity, calculated from these selected events,
  reaches a $3 \sigma$ level, with a $\Delta \chi^2$ of 9.}
 \end{abstract}

 \maketitle

\section{Introduction}

Neutrino oscillations provide us with the first glimpse of physics beyond the
standard model. They explain the observed deficits in the solar  
and the atmospheric neutrino fluxes.
The mass-squared differences needed to solve the solar neutrino problem
\cite{Bahcall:2004ut} and the atmospheric neutrino problem \cite{Fukuda:1998mi} are widely
different. Hence at least three neutrino mass eigenstates are required. This requirement fits nicely with the picture of three
active neutrino flavours, established by the invisible decay width
of $Z^0$ boson \cite{ALEPH:2005ab}. The three flavours mix to form three non-degenerate
mass eigenstates with masses $m_1, m_2$ and $m_3$. We get two independent mass-squared
differences $\Delta m^2_{21} = m_2^2 - m_1^2 = \Delta m^2_{\rm solar}$ and 
$\Delta m^2_{31} = m_3^2 - m_1^2 = \Delta m^2_{\rm atm}$. Solar and atmospheric
neutrino data indicate that $\Delta m^2_{\rm solar} \sim 0.03 \Delta m^2_{\rm atm}$.
Hence the third mass-squared difference $\Delta m^2_{32} = \Delta m^2_{31} -
\Delta m^2_{21}$ is approximately equal to $\Delta m^2_{31}$. The energy dependence
of the solar neutrino survival probability requires $\Delta m^2_{21}$ to be positive
but there is no experimental information on $\Delta m^2_{31}$. Thus two very different
patterns of neutrino masses are allowed: $m_1 < m_2 < m_3$ called normal hierarchy
(NH) and $m_3 < m_1 < m_2$ called the inverted hierarchy (IH). 
 
The unitary mixing matrix, relating the flavour eigenstates to the mass eigenstates,
is parameterized by three angles $\theta_{12}, \theta_{13}$ and $\theta_{23}$ and one CP violating phase 
$\delta_{CP}$. In the three flavour oscillation framework, it was shown that the solar
neutrino survival probability depends on the mixing angles $\theta_{12}$ and $\theta_{13}$
and the atmospheric $\nu_\mu$ survival probability depends on $\theta_{13}$ and $\theta_{23}$ \cite{Kuo:1986sk,Pantaleone:1993di}.
CHOOZ reactor neutrino experiment set a strong upper limit on $\theta_{13}$ of 
$\sin^2 2 \theta_{13} \leq 0.1$ \cite{Apollonio:1997xe,Narayan:1997mk}. In the limit of $\theta_{13} \to 0$, the solar neutrino
oscillations become effective two flavour oscillations, controlled by $\Delta m^2_{21}$ and
$\theta_{12}$. Similarly the atmospheric neutrino oscillations are also effective two 
flavour oscillations, controlled by $\Delta m^2_{31}$ and $\theta_{23}$. 
A systematic program of experiments with both natural \cite{Abdurashitov:2002nt,Hampel:1998xg,Fukuda:1998fd,Ahmad:2001an} and man made sources 
\cite{Araki:2004mb,Michael:2006rx,Abe:2011fz,An:2013zwz,Ahn:2012nd,Abe:2013hdq,Abe:2014ugx} have led to a wealth of
data on neutrino oscillation parameters. A three flavour oscillation fit to all the data
gives the following values for these parameters, as in table~\ref{Tab0}.


\begin{table}[H]				
  \centering
\begin{tabular}{|c|c|c|}
 \hline {\bf $\nu$-oscillation parameters}		&	{\bf Best fit values}	&	{\bf 3$\sigma$-range}	\\
 \hline
 \hline $\Delta m^2_{21}$ in 10$^{-5}$eV$^2$		&	7.60			&	7.11-8.18		\\
 \hline $\Delta m^2_{31}$ (NH) in 10$^{-3}$eV$^2$	&	2.48			&	2.30-2.65		\\
 \hline $\Delta m^2_{31}$ (IH) in 10$^{-3}$eV$^2$	&	2.38			&	2.20-2.54		\\
 \hline $\sin^2\theta_{12}$				&	0.323			&	0.278-0.375		\\
 \hline $\sin^2\theta_{23}$ (NH)			&	0.567			&	0.392-0.643		\\
 \hline $\sin^2\theta_{23}$ (IH)			&	0.573			&	0.403-0.640		\\
 \hline $\sin^2\theta_{13}$ (NH)			&	0.0234			&	0.0177-0.0294		\\
 \hline $\sin^2\theta_{13}$ (IH)			&	0.0240			&	0.0183-0.0297		\\
 \hline $\delta_{\rm CP}/\pi$ (NH) 			&	1.34			&	0.0-2.0			\\
 \hline $\delta_{\rm CP}/\pi$ (IH)			&	1.48			&	0.0-2.0			\\
 \hline
\end{tabular}
\caption{Best fit results and the 3$\sigma$-range of the global 3$\nu$ oscillations, as from the reference \cite{Forero:2014bxa} }
\label{Tab0}
\end{table}

The following questions still remain unanswered in the neutrino oscillation studies:
\begin{itemize}
 \item What is the pattern of neutrino masses? Is the true hierarchy normal or
 inverted?
 \item What is the octant of the angle $\theta_{23}$? Is it $< 45^\circ$ or $>45^\circ$?
 \item Most importantly, is there CP violation in the neutrino sector?
\end{itemize}
A number of experiments are currently running \cite{Abe:2013hdq,Abe:2014ugx,Ayres:2004js} or being planned 
\cite{Athar:2006yb,Aartsen:2014oha,Abe:2015zbg,Adams:2013qkq,An:2015jdp,seoppt} to address these issues. Among the
current experiments, NO$\nu$A can determine the hierarchy for the following two favorable combinations:
(i) The hierarchy is normal and $\delta_{CP}$ is in the lower half-plane or (ii) The hierarchy is inverted
and $\delta_{CP}$ is in the upper half-plane. If nature chooses one of the other two combinations, then
NO$\nu$A has no hierarchy sensitivity \cite{Huber:2009cw,Prakash:2012az}. Present data shows a slight preference
for normal hierarchy and $\delta_{CP}$ in the lower half-plane \cite{Abe:2013hdq,sanchezppt}.

When a neutrino passes through a medium, its propagation gets modified due to the coherent forward scattering. 
All three flavours undergo this scattering due to neutral current (NC) interactions whereas only $\nu_e$ has an additional scatteting 
amplitude due to charged current (CC) scattering off electrons \cite{Wolfenstein:1977ue,Mikheev:1986gs}. The scattering amplitudes give rise to potential
terms in the evolution equation. Since the NC interactions of all flavours are identical, the NC potential 
term does not lead to any modification of the oscillation probabilities. The CC potential term can lead to 
observable changes in the oscillation and survival probabilities. For the muon neutrino survival probability,
$P_{\mu \mu}$, a large change is possible only if $\Delta m^2_{31}$ is positive, which corresponds to normal
hierarchy (NH). For $\Delta m^2_{31}$ negative, called inverted hierarchy (IH), the change in $P_{\mu \mu}$
is negligible. For anti-neutrino the situation is reversed. The changes in the muon neutrino 
(or anti-neutrino) survival probability are significant when the following two conditions
are satisfied \cite{Gandhi:2004bj}:
\begin{eqnarray}
 \Delta m^2_{31} \cos 2 \theta_{13} & \sim & \pm 2 E V_{CC},   \label{resonance} \\
 \sin^2\left( 1.27 \frac{\Delta m^2_{31} \sin 2 \theta_{13} L}{E} \right) & \sim & 1, \label{peakprob} 
\end{eqnarray}
where $E$ is the energy of the neutrino, $V_{CC}$ is the potential due to CC scattering and $L$ is the pathlength
of the neutrino. The Wolfenstein matter term $A = 2 E V_{CC}$ (in eV$^2$) is given by $0.76 \times 10^{-4}
\rho \ ({\rm in \ g/cc}) \ E \ ({\rm in \ GeV})$, where $\rho$ is the density of the matter through which the
neutrino propagates. For $\Delta m^2_{31} \approx 2.5 \times 10^{-3}$ eV$^2$, eq.~(\ref{resonance}) is satisfied
for $\rho E \approx 33$. For the density $5$ gm/cc of earth's mantle, the corresponding energy is
$E \approx 7$ GeV. Substituting this in eq.~(\ref{peakprob}), we obtain a pathlength $L$ of the order of a few thousand km.
A large majority of upgoing atmospheric neutrinos pass only through earth's mantle hence the conditions mentioned
above are the most relevant.
Thus, we find that there is a broad range of energies around $7$ GeV and a broad range of pathlengths of
a few thousand kilometers for which there is an observable change in the muon neutrino survival probability
\cite{Gandhi:2004bj}. However, these changes can be measured and hierarchy can be determined only if the detector 
has good energy and direction resolutions \cite{Petcov:2005rv}.

\begin{figure}[H]
\centering
\includegraphics[width=1.\textwidth]{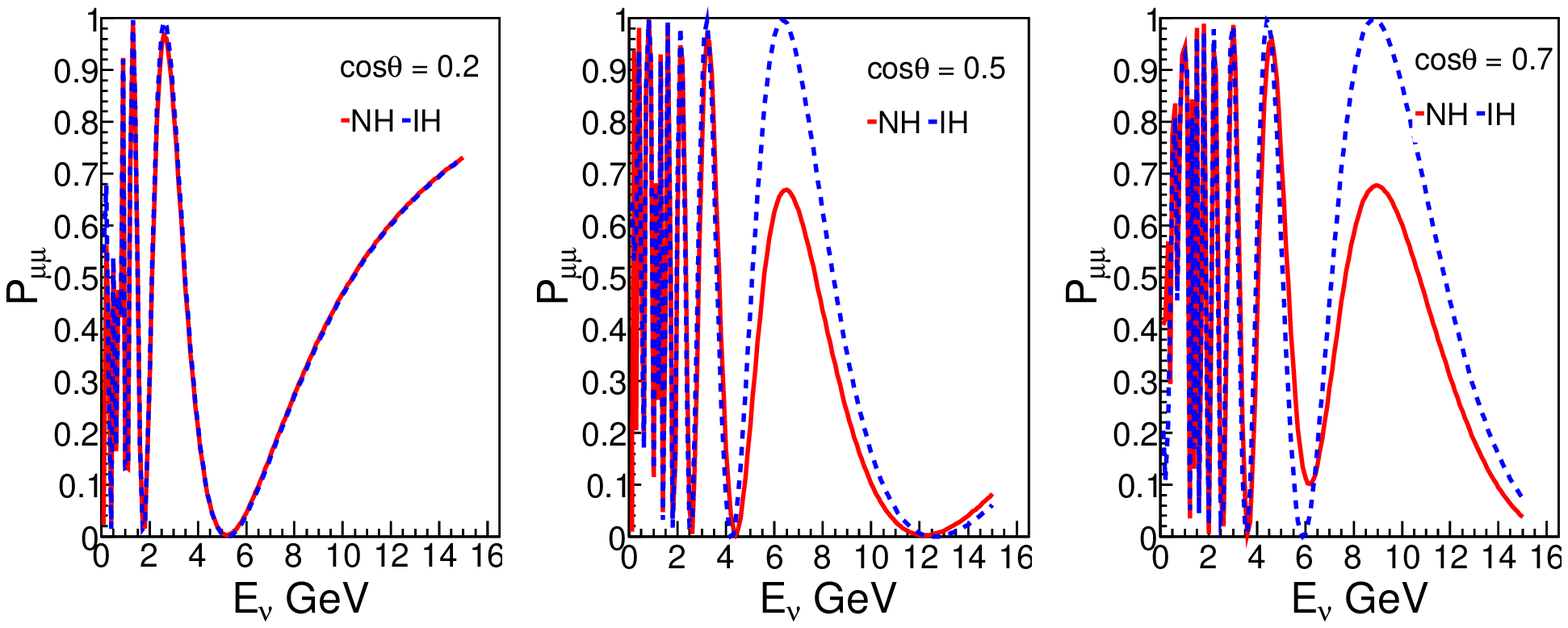}
\includegraphics[width=1.\textwidth]{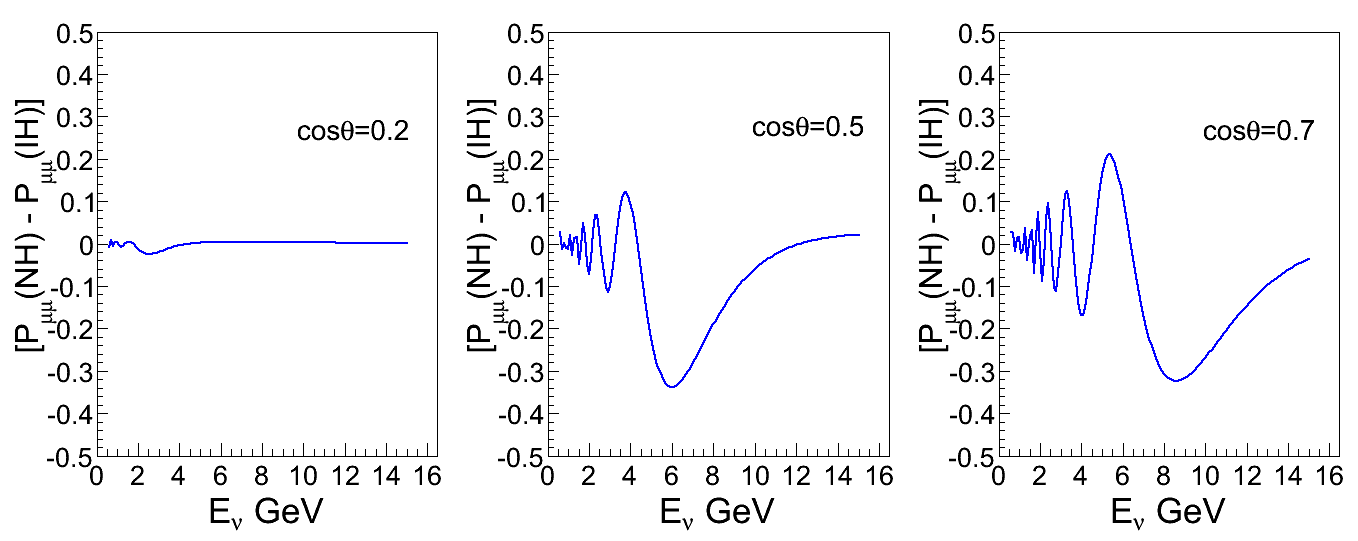}
\caption{Oscillation probability plot P$_{\mu \mu}$ for neutrinos at different zenith angles for either hierarchies (top panel). Difference between the values of P$_{\mu \mu}$ in the NH and the IH conditions (bottom panel). }
\label{Fig0}

\end{figure}

The plots of the muon neutrino survival probability $P_{\mu \mu}$ for atmospheric neutrinos,
as a function of neutrino energy, are shown in figure~\ref{Fig0} for various different values of 
$\cos \theta_z$, where $\theta_z$ is the zenith angle.
From these plots, we note that the signature for the neutrino mass hierarchy is most prominent
in the energy range $E_\nu >$ 4 GeV and for $\cos \theta_z > 0.5$. Hence, for the purpose
of hierarchy determination, $\nu_\mu$CC events in the {\it vertical cone} (i.e. with 
$|\cos \theta_z| > 0.5$) with $E_\nu >$ 4 GeV should be considered as the signal events 
and all other events should be termed background. It is imperative to develop a procedure by 
which it is possible to select the signal events with high efficiency and purity. In this
report, we develop such a procedure based on artificial neural network. 

\section{INO and ICAL}

India-based Neutrino Observatory (INO) \cite{Athar:2006yb} is an upcoming experimental facility
which can house a number of experiments requiring low cosmic ray backgrounds. A major component
of the experimental program at INO is the magnetised Iron Calorimeter (ICAL) neutrino detector. 
It will study neutrino interactions of various types, with the atmospheric neutrinos as the source. 
ICAL will be constructed in 3 modules, each of which contains 151 horizontal iron layers. The iron layers
are 5.6 cm thick and they are interspersed with resistive plate chambers (RPCs) 
\cite{Cardarelli:2007eb,Riegler:2002vg,Lippmann:2003uaa}. The total mass of the detector is made very large (more than
50 kilotons) to obtain a large sample of neutrino interaction. The area of each RPC is approximately
$2 {\rm m} \times 2 {\rm m}$ and the total number of RPCs in the detector is $\sim$30,000. 

When a neutrino interacts with an iron nucleus, it produces
a set of charged particles. These charged particles pass through one or more RPCs, depending on
the type of the particle and its energy. Whenever a charged particle passes through an
RPC, it produces a hit. These hits are our primary observables. 
The layer number of RPC gives the z-coordinate of the hit. The x and y-coordinates are
given by the copper-strips of the pick-up panels which are orthogonally oriented at the top and 
the bottom of the RPCs  \cite{2008arXiv0810.4693B}. 

ICAL has a good ability to identify muon tracks, from the pattern of hits in successive layers,
and determine the energy and the direction of the muons with good precision \cite{Chatterjee:2014vta}. In the 
present case, our {\bf signal} events are $\nu_\mu$CC events in the energy range of approx. 4-10 GeV, which
are in the vertical cone. These events are expected to have long muon tracks passing through many layers.
The background events come from sources: (a) low energy $\nu_\mu$CC interactions, (b) $\nu_\mu$CC interactions where
neutrino direction is close to the horizontal, (c) $\nu_e$CC events and a small number of $\nu_\tau$CC events and
finally (d) NC events. A large number of these background events do not give clear muon tracks. 
We aim to select a set of events which is highly rich in signal events. These events will have long muon tracks which can be recognized in a straight forward manner. We will utilize this fact to design criteria to separate the signal events from the background. 
 
We exploit the energy and the direction information of the neutrinos given to us by Nuance, in order to devise/develop the selection criteria. The selection criteria, whose developement is described in detail in the next section, depend only on the visible characteristics of the $\nu$-events in the detector, i.e. the output parameters given by GEANT4 simulation. The neural network is trained with a selected set of events. This trained network is then applied on any random set of events. It assigns a probability to the event, denoting how close it is to a perfect signal event. The signal-like events are finally chosen based on this value. The choice of the cut is such that the signal selection efficiency and the signal purity are significantly high.

In the previous section, we argued that events with $E_\nu >$ 4 GeV and $\cos \theta_z >$ 0.5 have the best
hierarchy sensitivity. However, neutrinos over a very broad range of energy and over the full zenith angle range
interact in the ICAL and produce observable events. Our job here is to develop a procedure to distinguish between
the events with good hierarchy sensitivity and those without. Hence neutrino events over the full detectable range
of energy must be simulated. 
Five different sets of data, each equivalent to 500 years of ICAL run, are generated using NUANCE in the energy range E$_\nu$= \{0.1,100\}GeV. Each set consists of 
all types of interactions of all three neutrino flavours. The first set is generated with the assumption of no neutrino
oscillations (NOOSC). The next three sets are generated assuming neutrino oscillations with normal hierarchy, with 3 different seeds. The final set is generated assuming neutrino oscillations with inverted hierarchy. The generated events are then propagated in ICAL using a Geant4 simulation of the detector. The pattern of hits thus generated are used to first identify the muon track and then reconstruct its energy $E_\mu$ and the cosine of its zenith angle $\cos \theta_\mu$ \cite{Redij}. In computing the oscillation probabilities, the following values of neutrino parameters were
used: $\Delta{{m}_{21}}^2 = 7.5\times 10^{-5}$ eV$^2$, $|\Delta{{m}_{eff}}^2| = 2.47\times 10^{-3}$ eV$^2$ (i.e. $\Delta{{m}_{31}}^2 ({\rm NH}) = 2.51\times 10^{-3}$ eV$^2$, $\Delta{{m}_{31}}^2 ({\rm IH}) = - 2.43\times 10^{-3}$ eV$^2$),  $\sin^2\theta_{12}$ = 0.31, $\sin^2{2\theta_{13}}$ = 0.09, $\sin^2{\theta_{23}}$ = 0.5 and $\delta_{CP}$ = 0.

\section{Effective Selection Parameters}

We first select only those events with hits in more than five layers $(L > 5)$. 
This lower limit on the number of layers is chosen to optimize the reconstruction efficiency
of the muon tracks \cite{Bhattacharya:2014tha}. This cut also has the advantage of eliminating most of the background
due to the non-$\nu_\mu$CC events. About 90\% of the events selected after this cut are $\nu_\mu$CC events \cite{Ajmi:2015qda}.
We want the neural network to select signal events with high efficiency and good purity. We need to choose appropriate 
input variables for the neural network to achieve this aim. We consider a number of such variables and study their
ability to distinguish between signal and background among the $\nu_\mu$CC events passing the $L>5$ cut.
For this study, we divide these events into four subsets based on the neutrino energy and direction, using the 
information from the event generator.
\begin{enumerate}
 \item  {Signal events}: $E_\nu$: 4-100 GeV and $|\cos \theta_z| > 0.5$,
 \item  {High energy horizontal events}: $E_\nu$: 4-100 GeV and $|\cos \theta_z| \leq 0.5$,
\item  {Low energy vertical events}: $E_\nu$: 0.1-4 GeV and $|\cos \theta_z| > 0.5$,
\item  {Low energy horizontal events}: $E_\nu$: 0.1-4 GeV and $|\cos \theta_z| \leq 0.5$.
\end{enumerate} 
We have checked that these variables discriminate against non-$\nu_\mu$CC background very effectively. 

 \subsection{ Hits} 
\begin{figure}[H]	
 \centering
 \setlength\fboxsep{0pt}
  
 {\includegraphics[width=1.\textwidth]{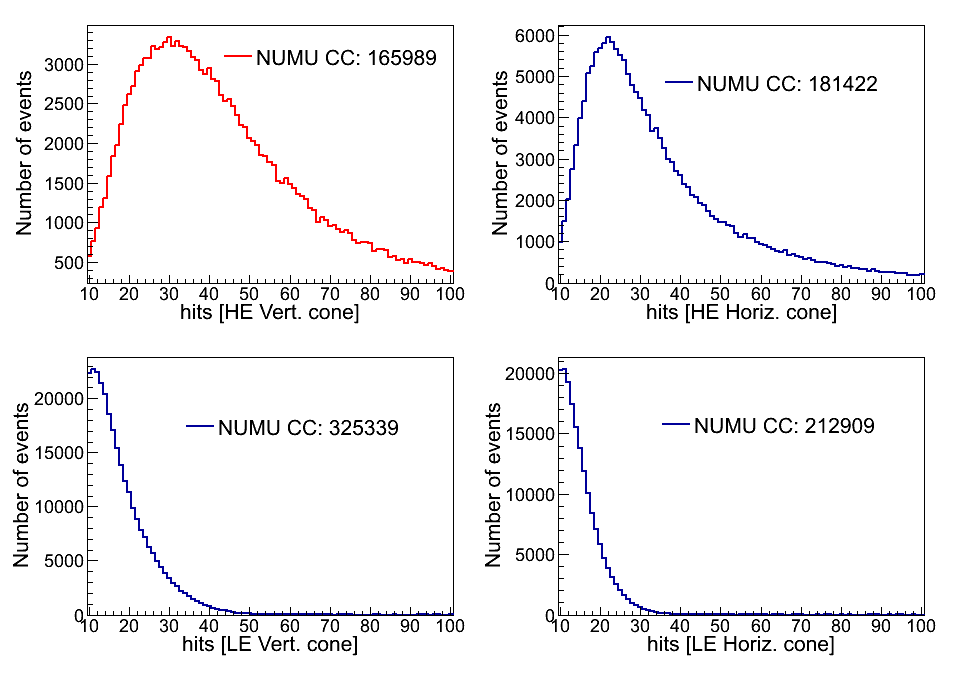} }
 \caption{Hits Distributions for $\nu_\mu$CC events with L$>$5 for the NOOSC dataset. The top left plot shows the distribution for the signal events (red).  }
\label{Fig1}
\end{figure}

 Low energy neutrino events give less number of hits compared to the high energy events. Hence, the number of hits is a measure of the energy of the neutrino as illustrated in figure~(\ref{Fig1}). This variable is quite effective in distinguishing high energy events from low energy events but not for distinguishing vertical events from horizontal events. 
 Figure~\ref{Fig1} shows that the signal-like events (top-left) give more hits than the low energy neutrino events (bottom row). The high energy horizontal events too give comparatively lower number of hits, if closely observed. This is due to the fact that the particles effectively travel through larger lengths of iron in the horizontal direction.
 \subsection{ Layers}
This parameter refers to the number of layers in ICAL, which has received one or more hits in an event. 
 The high energy vertical neutrino events give hits in more number of layers than the low energy/horizontal events. 
So, the vertical $\nu_\mu$CC events containing high energy muon tracks give hits in a larger number of layers than the other event types. 
 
 \subsection{Maximum horizontal spread of an event (maxdist)}
 
\begin{figure}[H]	
 \centering
 \setlength\fboxsep{0pt}
 {\includegraphics[width=1.\textwidth]{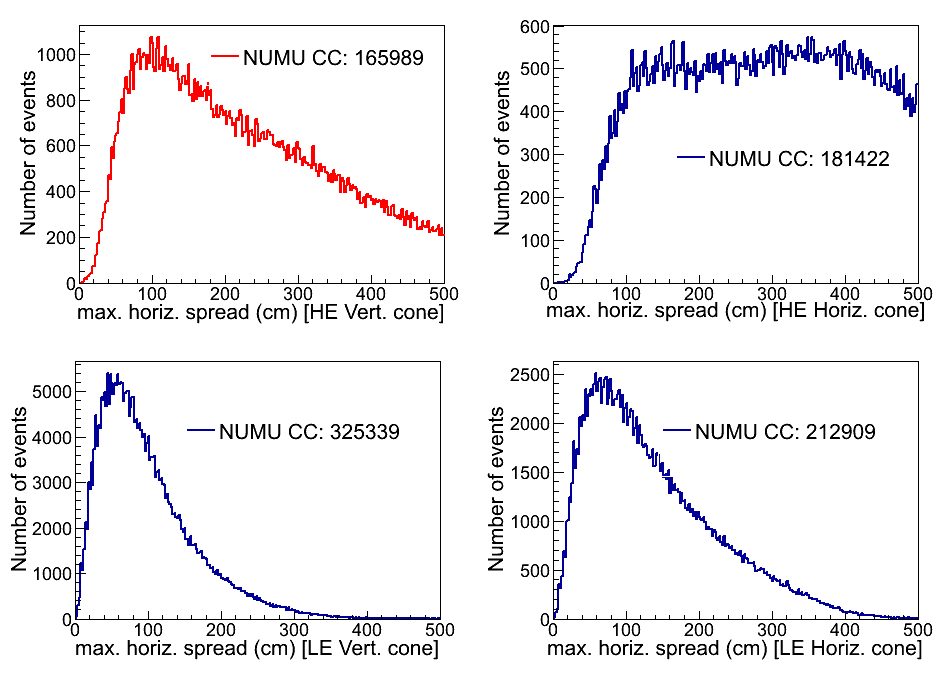} }
 \caption{Maxdist distributions for $\nu_\mu$CC events with  L$>$5 for the NOOSC dataset. The top left plot shows the distribution for the signal events (red).   }
\label{Fig3}
\end{figure}

 The energetic but near horizontal muons have a larger spread on the horizontal plane than the vertical (or near-vertical) events.
 The horizontal spread between a pair of hits is given by $D = \sqrt{((x_2\ -x_1)^2\ +\ (y_2\ -y_1)^2)}$. 
 We calculate $D$ for every pair of hits in an event and define its maxdist to be the maximum value of $D$ \cite{Ajmi:2015qda}.
 The maxdist is quite large in case of high energy horizontal events and is moderate for 
 the other three types of events, as can be seen in figure~\ref{Fig3}.
 
 



 \subsection{ Singlets}
 A high energy vertical muon passes through a number of layers. Any hadrons produced in the same event will pass through a much
 smaller number of layers. Therefore, in a high energy vertical event, there will be only hits due to the muon tracks after the 
 initial few layers. For these later layers, we expect one or two hits in a layer.
 Almost all signal events must contain one or more layers with a single hit. The passage of a muon through an
 RPC can produce a hit in a single strip or hits in two adjacent strips. Therefore, we define a layer with a single hit
 to be one where there is only one hit or one where there are two hits in adjacent strips.
 \emph{Singlets} is the number of layers in an event that contain a single hit. A signal event is expected to contain more singlets than the low energy or the horizontal $\nu_\mu$CC events. 
 
%
 \subsection{ Triplets}
 This is an extension of the previous parameter. 
 \emph{Triplets} is the number of 3 consecutive layers with single hits in an event. A signal event with a long muon track is 
 expected to contain at least one such triplet. This variable gives more weightage to events with longer muon track with many consecutive single hit layers. For example, an event with five consecutive single hit layers has {\it three} triplets. 
 
   \begin{figure}[H]	
 \centering
 \setlength\fboxsep{0pt}
 {\includegraphics[width=1.\textwidth]{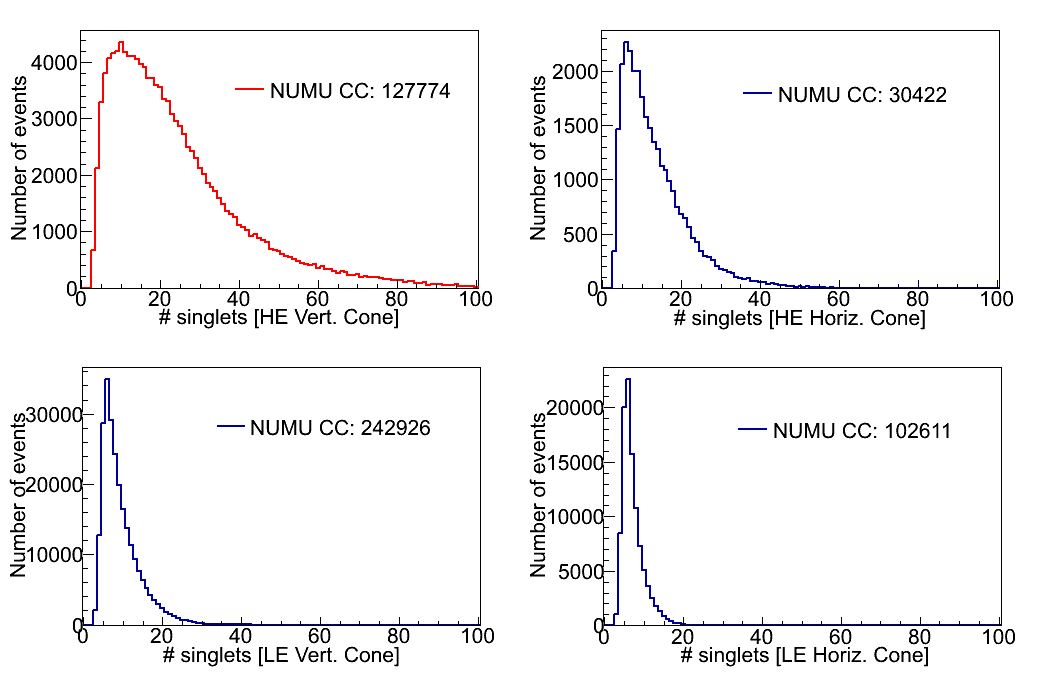} }
 \caption{Distribution of the single-hit layers in the $\nu_\mu$CC events (L$>$5), combined with maxdist/layers$\leq$10 and \# possible triplets$>$0, for the NOOSC dataset. The top left plot shows the distribution for the signal events (red).   }
\label{Fig5}
\end{figure}

Figure~\ref{Fig5} shows an example of the efficacy of the listed parameters in selecting the 
signal events and in discriminating against the background events. In this figure,
the singlet distribution is plotted for those satisfying the simple cuts: number of triplets non-zero
and the ratio maxdist/layers $\leq$ 10. These distributions show that a cut of number of singlets 
$\geq$ 10 retains most of the signal events while rejecting a very large fraction of background events.

 \subsection{Summarizing the effects of the selection parameters}

The above subsections all lead to the following inference:
\begin{itemize}
 \item Hits or layers can distinguish the low energy from the high energy range $\nu$ events.
 \item Maxdist distinguishes the horizontal high energy events from the rest.
 \item The high energy vertical $\nu_\mu$CC events contain significantly larger number of singlets than the low energy/ horizontal events.
 \item The hits-pattern across the layers in case of the high energy vertical $\nu_\mu$CC events form more number of triplets than the the other three categories of  $\nu_\mu$CC events considered.
\end{itemize}

\section{Choice of the analysis tool}
The parameters discussed in the above section indicate a reliable way to select our required signal events. A selection based on neural network techniques, with these parameters as inputs, can select the signal events efficiently. So, we employ tools for multi variate analysis (TMVA), a package integrated in ROOT, for our signal selection \cite{Hocker:2007ht}.

There are a number of applicable methods that involve multivariate analysis. Methods like BDT (Boosted Decision Trees), MLP (Multi-Layer Perceptron), LikelihoodPCA (Likelihood method using the input variables after Principal Component Analysis), HMatrix (method involving the inverse of the covariance matrix; a covariance matrix is devised based on the input variables for both, the signal and the background), and TMlpANN (a ROOT \cite{Brun:2000es} class TMultiLayerPerceptron) are all competitive. We did a preliminary analysis using each of these methods and compared the results. It was found that the event sample
selected by TMlpANN had the best signal efficiency and purity. In addition, the time taken for the method to learn the discrimination
and apply it to an event sample was also the least. In view of this the analysis was done using TMlpANN. Detailed 
optimization has shown that the stochastic learning method with three nodal steps and three hundred iterations gave the best performance.
We applied the method on a training set of 20000 events with the signal to background ratio being the same as that in
the actual data. The size of the training set was chosen after detailed optimization. The training and testing sets,
post optimization, were taken from the NOOSC data set.

During the training, the selection parameters, discussed in section 3, are given as inputs to
the adaptive neural network (ANN). The ANN combines these inputs in various weights to perform a number
of intermediate calculations. Over a number of iterations, the trained ANN optimizes
the weights and finally learns to assign a number to an event. This number is the probability of the event being a signal event.
After the training, the ANN is fed the events from the NH data sets and the 
IH data set. The signal events are selected from each data set based on a cut on the probability
assigned by the ANN (called ANNcut).

\begin{figure}[H]	
 \centering
 \setlength\fboxsep{0pt}
 {\includegraphics[width=1.\textwidth]{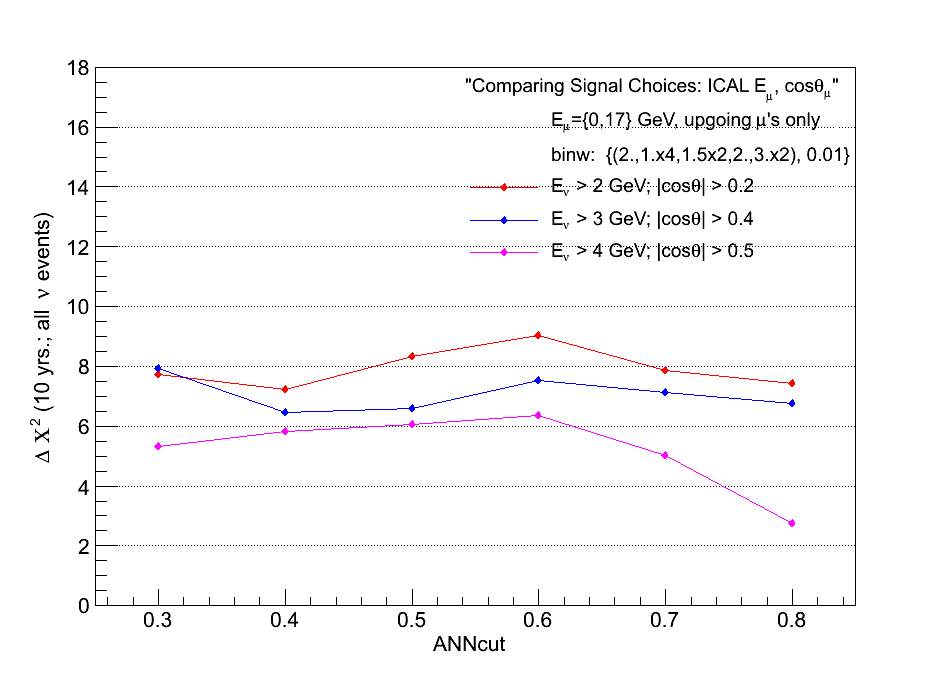} }
 \caption{Hierarchy discrimination sensitivity for different signal definitions }
\label{figcomp}
\end{figure}

In our initial analysis, we defined the signal events to be $\nu_\mu$CC events with $E_\nu > 4$ GeV
and $|\cos \theta_z| > 0.5$. But some hierarchy discrimination is present in events with lower energy
and smaller $|\cos \theta_z|$. So we systematically lowered the minimum values of $E_\nu$ and 
$\cos \theta_z$ and found that the best hierarchy discrimination sensitivity is obtained with $E_\nu > 2$ GeV and
$|\cos \theta_z| > 0.2$, as shown in figure~\ref{figcomp}. 
The details of the calculation of the hierarchy discrimination sensitivity are given in the next section.

\section{Calculation of Mass Hierarchy $<\Delta\chi^2>$}
Our redefined signal consists of the $\nu_\mu$CC events with E$_\nu$ = \{2,100\}GeV and $|cos\theta|>$ 0.2. Therefore, the background comprises of all the rest of the $\nu_\mu$CC events as well as all non-$\nu_\mu$CC events (i.e., all NCs, $\nu_e$CC and $\nu_\tau$CC).
If we impose the cut $L>5$ on the 500 year dataset with IH oscillations, the remaining data sample has the composition shown in table~\ref{Tab8}. The numbers for NH oscillations are similar.

 \begin{table}[H]					
 \centering
\begin{tabular}{|c|c|}
  \hline	 Total	&	 Total		\\
 	{ signal events}& { bkg. events}	\\
 \hline
	     $\sim$350,000		&$\sim$400,000		\\
 \hline
 \end{tabular}
 \caption {Counts of the events after a cut of Layers$>$5, in a 500 years data sample.}
\label{Tab8}
\end{table} 
	 
For the selected signal-like events, the muon energy $E_\mu$ and its direction $\cos \theta_\mu$ are reconstructed.
The events are sorted into bins of the reconstructed $E_\mu$ and $\cos \theta_\mu$. A better angular resolution
leads to a better hierarchy discrimination in atmospheric neutrino experiments \cite{Petcov:2005rv,Gandhi:2007td}.
Hence the bin width in $\cos \theta_\mu$ needs to be as small as possible. Here it is chosen to be $0.01$, which is
the $\cos \theta_\mu$ resolution of the ICAL \cite{Chatterjee:2014vta}.
The down going events undergo no oscillation and hence there will not be any signature of matter effect in them.
It is present only in the up going events. Hence in computing the $\Delta \chi^2$ for hierarchy signal, we will consider
only the up going events, {\it i.e.} events with $\cos \theta_\mu \in [0,1]$. This has the advantage of eliminating the 
contribution of the fluctuations in the down going events to the $\chi^2$.

 \begin{figure}[H]	 					
 \centering
  {\includegraphics[width=1.\textwidth]{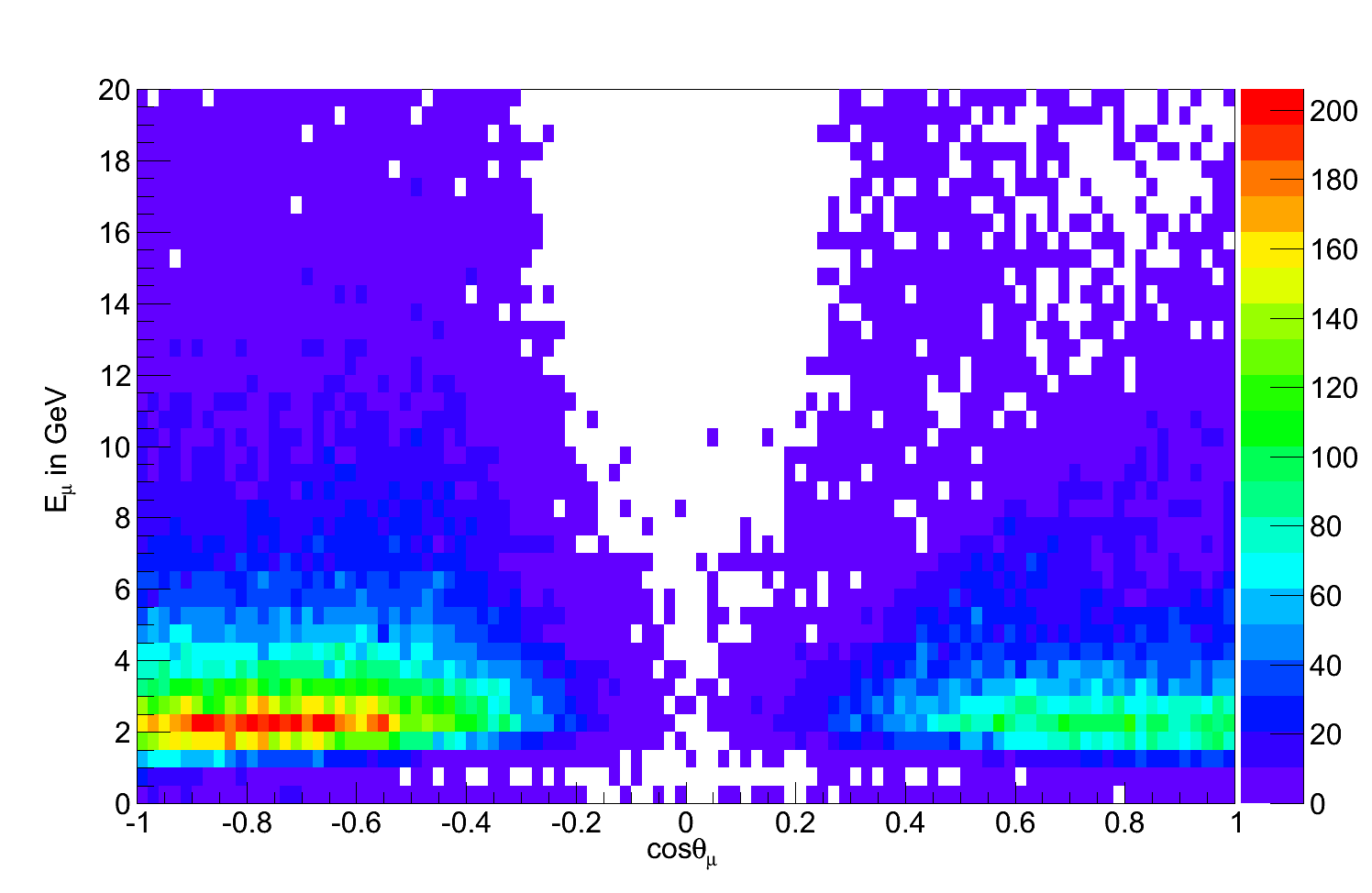} }
\vskip -2ex    \caption{The distribution of muon energy and direction (given by NUANCE) of the events surviving 
the cut on the neural network probability = 0.7.  }
 \label{Fig8b}
\end{figure}

Regarding the binning in muon energy, different schemes were tested. 
We restricted the range of muon energies to be (0,17) GeV. Since events with very high energy
muons are rather small in number, their contribution to hierarchy discrimination is very small.
Also, the smaller event numbers have large fluctuations which lead to spurious contribution to
hierarchy sensitivity. Therefore a binning scheme with uniform energy bins is not preferred.
We have verified that the results are much better for a scheme with differential energy bins
compared to a scheme with uniform energy bins.
The $L>5$ cut as well as the selection based on ANN strongly discriminate against events with E$_\mu<$ 1 GeV. 
This  can be seen in figure~\ref{Fig8b}. Therefore, our lowest energy bin is chosen to be (0, 2) GeV, so that 
the number of events in this bin are substantial.
We found the following 10 $E_\mu$ bins to be optimal: (0, 2), (2, 3), (3, 4), (4, 5), (5, 6), (6, 7.5), (7.5, 9),
(9, 11), (11, 14) and (14, 17) GeV. The binning is done separately for $\mu^-$ and $\mu^+$ events. 
Differential binning schemes are also tried out for $\cos \theta_\mu$. But results of the uniform binning
scheme with $0.01$ bin width are always better.

\begin{figure}[H]	 					
 \centering
  {\includegraphics[width=1.\textwidth]{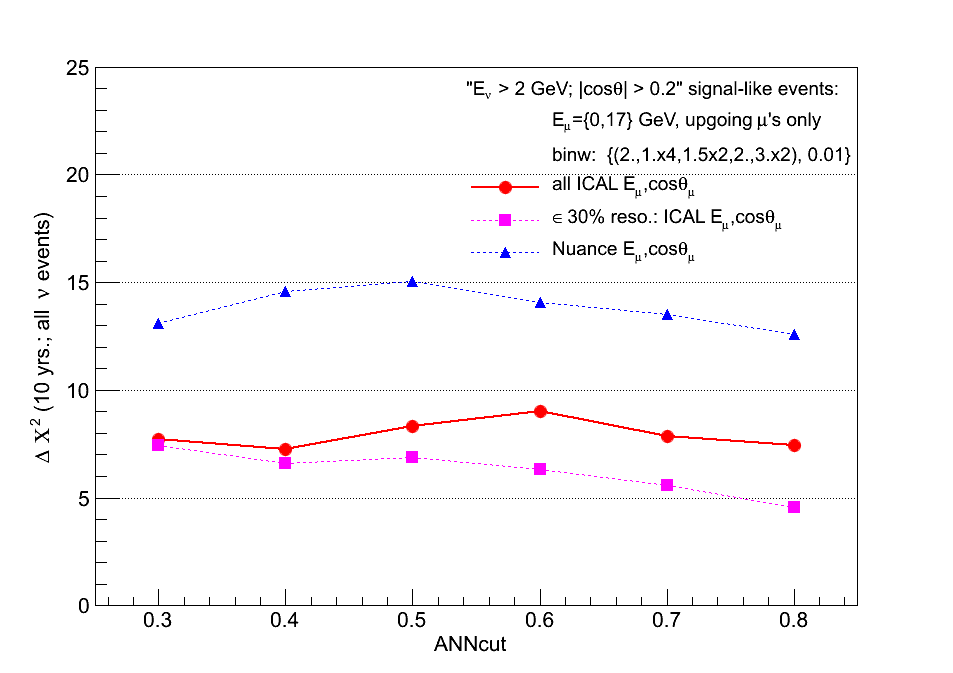} }
\vskip -2ex    \caption{Values of $<\Delta\chi^2>$ for 10 years, against varying cuts on the probability obtained from the neural network analysis, in the range E$_\mu$ = \{0, 17\} GeV and $\cos \theta >$ 0, using a differential binning scheme. 
The red circles correspond to the case where E$_\mu$ and $\cos\theta_\mu$ are obtained from the ICAL reconstruction code.
The magenta squares correspond to the case where E$_\mu$ is restricted to be within 30\% of its Nuance value.
The blue triangles show the result assuming an ideal momentum reconstruction. }
 \label{Fig9}
\end{figure}

For the hierarchy discrimination analysis the four data sets $NH1, NH2, NH3$ and $IH$ were used. Each of them was binned according to the scheme described above. From this binned data, 
         we compute 3 values of $\chi^2_{\rm true}$ as $\chi^2(NH1-NH2)$, $\chi^2(NH1-NH3)$ and $\chi^2(NH2-NH3)$, and take their average to obtain $<\chi^2_{true}>$. This is expected to be twice the number of bins.
         We also compute 3 values of $\chi^2_{\rm false}$ as $\chi^2(NH1-IH)$, $\chi^2(NH2-IH)$ and $\chi^2(NH3-IH)$, and take their average to obtain $<\chi^2_{false}>$.
	 From these we obtain $ <\Delta\chi^2>\ =\ <\chi^2_{false}>\ -\ <\chi^2_{true}>$.
We obtained a maximum $<\Delta \chi^2> = 9$ (assuming a 10 year run) for an ANN probability cut of 0.6. The variation in 
$<\Delta \chi^2>$ with the ANN probability is plotted in figure~\ref{Fig9}.

Figure~\ref{Fig9} contains two other plots which are put in for comparison. The blue curve in figure~\ref{Fig9} shows the hierarchy discrimination of an ideal detector, which gives the exact values of $E_\mu$ and $\cos \theta_\mu$. This sets the upper limit on the hierarchy discrimination of ICAL, when only $E_\mu$ and $\cos \theta_\mu$ 
are used as inputs. In addition, we considered the possibility that the fluctuations due to inefficiencies 
in the reconstruction are responsible for $<\Delta \chi^2>$.  To rule out this possibility, we considered only those 
events which satisfied the following two criteria: (i) The charge of the muon is correctly identified and 
(ii) The reconstructed $E_\mu$ is within $30\%$ of the NUANCE value. These restrictions give us a smaller event
sample. The $<\Delta \chi^2>$ from this subset is shown as the magenta curve in figure~\ref{Fig9}. For this restricted set, the maximum
$<\Delta \chi^2> = 7$, which is close to the value of $9$ obtained for the full sample. Therefore, the hierarchy
discrimination we obtained comes mostly from well reconstructed events.

\section{Discussion}

We have used a neural network to identify high energy $\nu_\mu$CC events in the vertical direction. The neural network is 
able to select such events with an efficiency $>70\%$. The purity of the selected samples is also $>70\%$. We have binned the selected signal
events in $E_\mu$ and $\cos \theta_\mu$ to compute the hierarchy sensitivity of ICAL. Since this sensitivity depends strongly
on the angular resolution, we chose a rather fine bin width of $0.01$ for $\cos \theta_\mu$. For binning $E_\mu$, we found
that a scheme with differential bin widths gives a much better sensitivity. Hence we chose a set of ten bins with finer bins
at lower energy and wider bins at higher energy. We also imposed an upper limit $E_\mu \leq 17$ GeV. 

\begin{figure}[H]	 					
 \centering
  {\includegraphics[width=1.\textwidth]{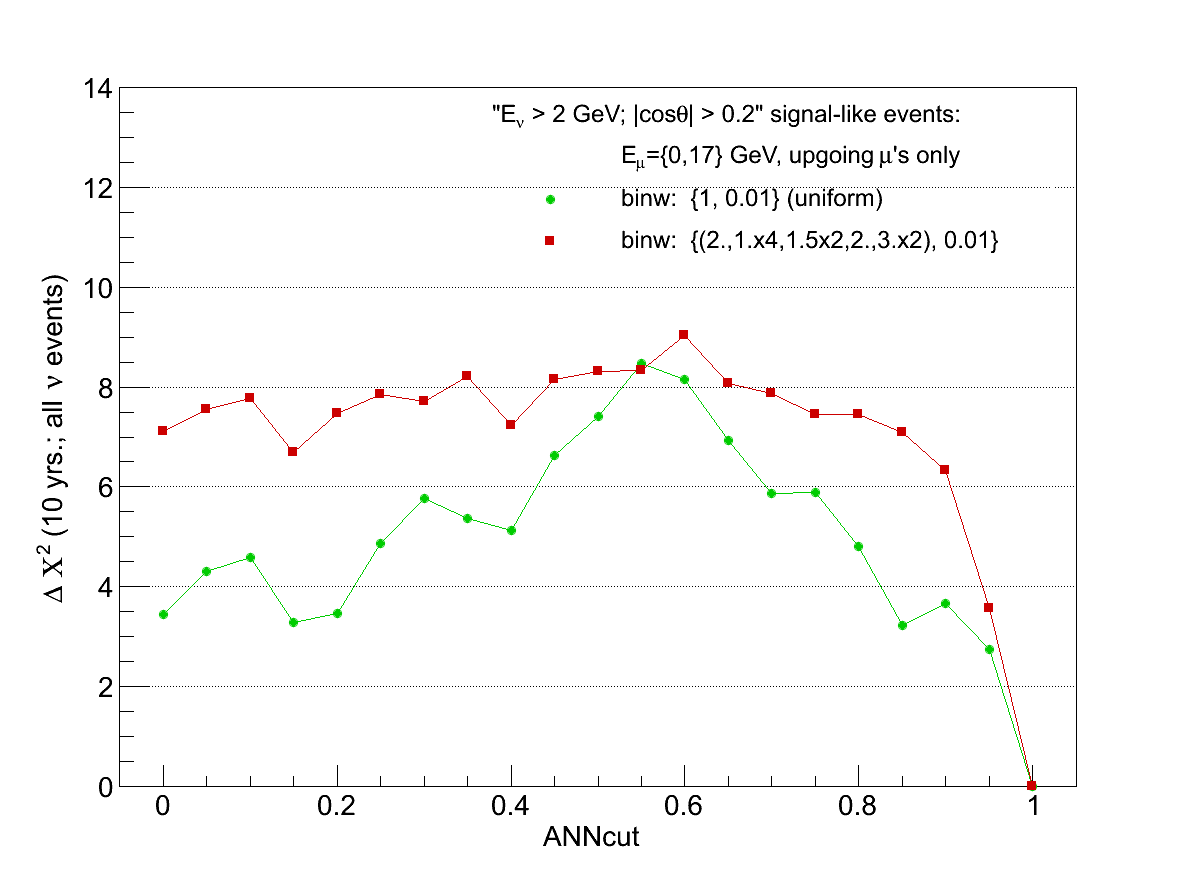} }
\vskip -2ex    \caption{Values of $<\Delta\chi^2>$ for 10 years against ANNcut, in the range E$_\mu$ = \{0, 17\} GeV and $\cos \theta >$0, using a uniform binning scheme and a differential binning scheme.  }
 \label{Fig10}
\end{figure}

From figure~\ref{Fig9}, we see that the $<\Delta\chi^2>$ has only a mild dependence on ANNcut. Hence we did a brief study on the role played by the ANN in signal selection. These results are summarised in figure~\ref{Fig10}, which plots $<\Delta\chi^2>$ vs. ANNcut, for uniform energy binning scheme and for differential energy binning scheme. The $<\Delta\chi^2>$  for the differential scheme is always larger than the uniform scheme. In case of the uniform scheme, signal selection through ANN plays an important role and improves $<\Delta\chi^2>$ from 3.5 to 8. On the other hand, the differential scheme inherently extracts the differences between the event spectra of NH and IH. For this scheme, the event selection due to ANN leads only to a small improvement in $<\Delta\chi^2>$ from 7 to 9.

For the sample we chose, we obtained a maximum $<\Delta \chi^2> = 9$, {\it i.e.} a $3 \sigma$ hierarchy discrimination.
We have calculated the increase in $<\Delta \chi^2>$ which can happen in the case of an ideal detector. It is found that 
with absolute energy reconstruction, very fine direction resolution and perfect charge identification, one can obtain a 
$<\Delta\chi^2> \sim 15$. Therefore improvement in the muon energy reconstruction and charge identification will certainly
improve the hierarchy sensitivity of ICAL.

\section{Acknowledgement}
We thank the members of the INO collaboration for all the support and cooperation we received in the course of this work.
In particular, we thank Jim Libby for a discussion on the use of neural networks.

 \bibliographystyle{jhep.bst}

 \bibliography{/home/ali/UmaS/reports/Aug2014/SML_bibtex_inspire}

\end{document}